\documentclass[a4paper]{article}
\usepackage[explicit]{titlesec}
\usepackage{hyphenat}
\usepackage{ragged2e}
\usepackage{textcomp}
\RaggedRight
\usepackage[utf8]{inputenc}
\usepackage{multicol}
\usepackage[T1]{fontenc}
\usepackage{amsmath, amsthm}
\usepackage{amssymb}
\usepackage{graphicx}
\usepackage[linesnumbered,boxed,commentsnumbered,noline]{algorithm2e}
\usepackage{hyperref}

\usepackage{soul}
\setul{.6pt}{.4pt}

\usepackage{geometry}
\usepackage{fancyhdr}
\usepackage[labelfont={footnotesize,bf} , textfont=footnotesize]{caption}
\makeatletter
\renewcommand\tagform@[1]{\maketag@@@ {\ignorespaces {\footnotesize{\textbf{Equation}}} #1.\unskip \@@italiccorr }}
\makeatother
\titleformat{\section}
  {\normalfont}{\thesection}{1em}{\MakeUppercase{\textbf{#1}}}
\titlespacing\section{0pt}{0pt}{-10pt}
\titlespacing\subsection{0pt}{0pt}{-8pt}

\makeatletter
\newcommand\sixteen{\@setfontsize\sixteen{17pt}{6}}
\renewcommand{\maketitle}{\bgroup\setlength{\parindent}{0pt}
\begin{flushleft}
\sixteen\bfseries \@title
\medskip
\end{flushleft}
\textit{\@author}
\egroup}
\makeatother

\usepackage[sorting=none]{biblatex}
\title{A stochastic geospatial epidemic model and simulation for COVID-19 using an event modulated Gillespie algorithm}

\author{Alexander Temerev $^a$, Liudmila Rozanova $^b$, Olivia Keiser $^c$, Janne Estill $^d$\\}

\begin{document}

\vspace*{.01 in}
\maketitle {$^a$ Institute of Global Health, University of Geneva, 9 Chemin des Mines, 1202, Geneva, Switzerland, alexander.temerev@unige.ch\\ $^b$  Institute of Global Health, University of Geneva, 9 Chemin des Mines, 1202, Geneva, Switzerland, liudmila.rozanova@unige.ch\\ $^c$ Institute of Global Health, University of Geneva, 9 Chemin des Mines, 1202, Geneva, Switzerland, olivia.keiser@unige.ch \\ $^d$ Institute of Global Health, University of Geneva, 9 Chemin des Mines, 1202, Geneva, Switzerland, janne.estill@unige.ch\\
Corresponding author: Alexander Temerev, alexander.temerev@unige.ch}
\vspace{.12 in}
\justifying

\section*{abstract}
  
We developed a model and a software package for stochastic simulations of transmission of COVID-19 and other similar infectious diseases, that takes into account contact network structures and geographical distribution of population density, detailed up to a level of location of individuals. Our analysis framework includes a surrogate model optimization process for quick fitting of the model’s parameters to the observed epidemic curves for cases, hospitalizations and deaths. This set of instruments (the model, the simulation code, and the optimizer) is a useful tool for policymakers and epidemic response teams who can use it to forecast epidemic development scenarios in local environments (on the scale from towns to large countries) and design optimal response strategies. The simulation code also includes a geospatial visualization subsystem, presenting detailed views of epidemic scenarios directly on population density maps. We used the developed framework to draw predictions for COVID-19 spreading in the canton of Geneva, Switzerland.

\section*{keywords} 
\textit{COVID-19, stochastic epidemic modeling, Gillespie algorithm, contact matrices, population density, epidemic simulation}
\vspace{.12 in}

\section*{introduction}

Numerous epidemic modeling and simulation toolkits have been developed or adapted for COVID-19, ranging from educational models to global-scale comprehensive frameworks \cite{van2011gleamviz}. Network representations are powerful tools that allow us to understand disease transmission in human populations while taking into account the structure of human interactions, mobility and contact patterns. Previous studies have paid attention to the influence of various network configurations (scale-free, random, small-world) \cite{pastor2015epidemic, small2005small, de2018fundamentals} on the spreading rate and value of the epidemic threshold. Some models also work with population mobility data, which is especially important in the context of imposing movement restrictions to reduce transmission \cite{chang2021mobility, van2011gleamviz}.
To improve the accuracy of epidemic models, it is necessary to account for population heterogeneity in terms of, for example, age, social groups and mobility patterns, as well as geographical clustering of infection spreading that can arise from higher contact rates in places with higher population density. Examples of models with a detailed representation of these factors include the age structured household model of Pellis et al. \cite{Pellis2020} and the two-level (global and local) mixing model proposed by Ball et al. \cite{Ball1997} which has further been expanded to account for network structure  \cite{Ball2008}. However, these models do not include distance metrics or account for differences in population density. In the context of analyzing and predicting the progression of the COVID-19 epidemic, the impact of various epidemic control and containment measures needs thorough consideration \cite{aleta2020modelling}. 
The importance of the spatial component in epidemic systems is being increasingly recognised. When models are used to assess spatially heterogeneous interventions, they need to be able to represent the location of hosts and the spatial pattern of transmission in enough detail. Including the spatial component to epidemic models has previously been seen as a complex technical addition that should only be considered when absolutely necessary \cite{riley2015five}. However, as the software and methods improve and spatial individual-level data becomes increasingly available, we may expect that the spatial component will take an increasingly important role in infectious disease modelling tools.
We therefore aimed to develop a geospatial network model, and calibrate it to the COVID-19 data  from the canton of Geneva, Switzerland, to test its applicability and validity.

\section*{Methods}

\subsection*{General assumptions}

We use a standard representation of the population as a network consisting of $N$ nodes (individuals) $\{i\}$ and a set of links $\{e_{ij}\}$, representing a contact between each node $i$ and $j$. We assume that the number of individuals in the network remains constant during the simulation \cite{pastor2015epidemic}.
In our case the network is weighted (all links $e_{ij}$ have a weight depending on the spatial distance between the nodes $i$ and $j$) and complete (i.e. a link $e_{ij}$  exists for any pair of nodes $i$ and $j$).
We represent the infection process in the network with  a stochastic SEIR model, so that all nodes are at any time point in one of the four states: $S$ (susceptible), $E$ (exposed in the latent period), $I$ (infectious) or $R$ (recovered/removed).
Stochastic infection process $S\to E$ occurs as a result of contact between a susceptible ($i\in S$) and infectious ($j \in I$) individual. Two other random processes also take place in the network in parallel: the transition of individuals from the exposed to infectious state $E \to I$ and the removal process $I \to R$.
The model assumes two types of contact events, which we call “local” and “global” contacts and are distinguished using the Euclidean distance matrix $M = \{m_{ij}\}$. Thus, we consider separately  contacts that happen in the neighbourhood of each individual’s residence, and those that take place during longer-distance mobility.
Local contacts are determined by a cutoff $r$ on the distance $m_{ij}$. This cutoff is the same for all nodes in the network. Global spreading is modelled as a separate process where each infectious individual can spontaneously transmit the infection to a random target selected from the entire population. The process is automatically normalized for population density, as the target will be more likely located in densely populated areas.
In our model, the following parameters affect the contact probability between two nodes:

\begin{enumerate}
\item The assignment of the nodes into one of 16 age groups. The frequency of contacts between the age groups is given as an asymmetric 16x16 matrix $A = \{a_{ij}\}$ [Appendix A].
\item The distance between the nodes. As only relative distances matter in our model, we use the Euclidean distance between locations on the map. We determine a fixed cutoff distance to distinguish between local and global contacts. 
\item The population density in a particular area. The contact probability for global contacts directly depends on the population density in the area of the contacting nodes – this is achieved with rejection sampling of global contacts over the entire simulated region.
\end{enumerate}

\subsection*{Epidemic model}

We represent the stochastic SEIR model as a continuous time 3-dimensional Markov chain $X=\{(S(t),E(t),I(t)):t\ge 0 \}$ that tracks the number of susceptible, exposed and infectious individuals at any time point. The number of removed individuals can be calculated as $R(t)=N-S(t)-E(t)-I(t)$.

The epidemic starts with $n_0$ infectious individuals; the rest of the population is assumed to consist of fully susceptible individuals; that is, the initial state of $X$ is $(S(0),E(0),I(0))=(N - n_0,0,n_0)$. In each moment of time the state space of $X$ is described by changing the state of individuals according to the rules shown in Table 1. All other types of contacts do not change the state of the system.

\begin{table}
\begin{center}
\begin{tabular}{ c|c|c|c } 
 \hline
 Transition & Type & Rate & State change \\ 
 \hline
 $S \to E$ & Local/global contagion & $\beta_{i}$ & $(s_t-1,e_t+1,i_t)$ \\
 $E \to I$ & Spontaneous & $\epsilon_{i}$ & $(s_t,e_t-1,i_t+1)$ \\ 
 $I \to R$ & Spontaneous & $\gamma_{i}$ & $(s_t,e_t,i_t-1)$ \\ 
  \hline
\end{tabular}
\caption{Transition rates between Markov chain states}
\end{center}
\end{table}




\subsection*{Simulation algorithm}

To simulate the infection process, we developed and applied a new variant of the event modulated Gillespie algorithm. Our implementation supports multiple epidemic control regimes and arbitrary functional forms and distributions of epidemic parameters. It can be used to validate theoretical models and it is particularly suitable for the simulation of epidemic dynamics in large regions such as entire countries.

The unmodified Gillespie algorithm is specified as follows. Consider $N$ Poisson processes with the rate of $\lambda_i$, $i\in[1,N]$ running in parallel. We denote the density of the event rate for the $i$th process by $\rho_i(\lambda_i)$. The renewal process is fully characterised by the probability density of inter-event times, and denoted as $\phi_i(\tau)$ for the $i$th process.

For the Poisson process with probability density of the event rate $\rho_i(\lambda_i)$, we have
\begin{equation}
\psi_i(\tau) = \int_0^{\infty}\rho_i(\lambda_i)e^{-\lambda_i \tau}d\lambda.
\notag
\end{equation}
A Poisson process with rate $\lambda_0$, i.e.,  $\psi_i(\tau) = \lambda_0 e^{-\lambda_0 \tau}$ is generated by $\rho_i(\lambda_i) = \delta (\lambda_i - \lambda_0)$, where $\delta$ is the delta function.  
In the unmodified Gillespie algorithm applied to epidemic spreading, all possible state transitions for nodes ($S\to E$, $E\to I$, $I\to R$) are Poisson processes with rates $i$. Multiple simultaneous Poisson processes can be merged into a single global Poisson process with the rate $\Lambda = \sum_{i=1}^N\lambda i$, where time intervals between events are distributed exponentially with the geometric mean $1/\Lambda$ — and for every event of this process we need to calculate what exactly is this event, and where it has happened (for that, we have to keep track of all susceptible nodes that can currently be immediately infected, all exposed nodes that can transition to the infectious state, and all infectious nodes that can be recovered — and sample the probabilities of these transitions accordingly). The original Gillespie algorithm was first invented to track chemical reactions and later adapted to model epidemic processes in well-mixed populations, where the only important output are the counts of entities in each state (e.g. molecules in chemical models, individuals in epidemic models). However, in network and geospatial models, it is important to know which node exactly is making the transition, which leads to computational difficulties.

In our event-modulated Gillespie algorithm, we cut off all non-productive (not leading to a change of states) activations, and use a global event queue (a priority queue, i.e. all events are always sorted by time) to automatically place all actual state transitions to the appropriate time points in the future. Therefore, we no longer need to merge all event rates $i$ into the global rate $\Lambda$ but instead work directly with the SEIR parameters $\beta$, $\epsilon$, and $\gamma$, using them as the rates of the corresponding Poisson processes and generating exponentially distributed  time intervals between events. 

When a node becomes infectious (either from the Exposed state after the incubation period, or by being one of the initially infected patients), the infection sequence starts. First, we sample the time interval until removal (recovery or death) as $-\ln (U)/\gamma$, where $U$ is a random value sampled from the uniform distribution $[0..1]$, and place a removal event in the priority queue. Then, we generate exposure events and their time offsets, continuing as long as the sum of time intervals between these events is less than the time from becoming infectious to the removal event. The exposure process is described as follows. 

We first determine whether the exposure event will be local or global, the probability of a global transmission being $q$, and the probability of a local transmission $1-q$. We then sample an age group for the target node from the discrete distribution taken from a row of the contact matrix corresponding to the source node’s age group. If the exposed contact is local, we select randomly a node $j$ that is located within the Euclidean distance $m\leq r$ from the index node $i$ and has an age group equal to the age group we sampled in the previous step. If the exposed contact is global, we select randomly any node $j$ over the entire simulated geographical region using rejection sampling (by generating random uniformly sampled Euclidean coordinates with 10m precision until we have a location that contains a node from the chosen age group), so that the target node is selected according to the population density distribution. Then, we sample the time offset for this exposure event as $-ln(U)/\beta$. If the time of the exposure is still less than the time of the recovery, we place the event in the queue and move on to the next exposure event  If the exposure event is already beyond the removal time, we cancel the last exposure event and stop the process.

The algorithm itself is outlined in (Algorithm 1), adapted from Kiss et al.  \cite{kiss2017mathematics}.

\IncMargin{2em}
\begin{algorithm}[H]
\DontPrintSemicolon
\SetKwInOut{Input}{input}
\SetKwInOut{Output}{output}
\Input{N: array of nodes with 2D coordinates (all in $S$ state); $\beta_i, \epsilon_i, \gamma_i$: SEIR parameters; $t_i$: regime change dates; $n_{start}$: initial infected}
\Output{$W$: array of nodes with 2D coordinates and states at time $t$.}
\BlankLine
$Q \leftarrow \varnothing$ \tcc*[r]{priority queue}
$I \leftarrow$ random\_sample($N$, $n_{start}$)\;
\For{$n \leftarrow I$}{
$e \leftarrow$ Infect(node = $n$, time = 0)\;
$Q$.enqueue($e$)
}
\While{$Q$ not empty}{
$U \leftarrow$ random(0, 1)\;
$e \leftarrow Q$.dequeue()\;
$\beta_t, \epsilon_t, \gamma_t \leftarrow$ regime\_params($e$.time)\;
\If{e is Expose}{
$e_I' \leftarrow$ Infect(node = $e$.node, time = $e$.time $+$ $\frac{-\log U}{\epsilon_t}$)\;
$Q$.enqueue($e_I'$)\;
$e$.node.state $\leftarrow E$\;
}
\If{e is Infect}{
$e_E' \leftarrow$ Expose(node = find\_contact($e$.node), time = $e$.time $+$ $\frac{-\log U}{\beta_t}$)\;
$e_R' \leftarrow$ Remove(node = $e$.node, time = $e$.time $+$ $\frac{-\log U}{\gamma_t}$)\;
\If{$e_E'$.time < $e_R'$.time}{
$Q$.enqueue($e_E'$)\;
}
$Q$.enqueue($e_R'$)\;
$e$.node.state $\leftarrow I$\;
}
\If{e is Remove}{
$e$.node.state $\leftarrow R$\;
}
}
\caption{Event-modulated Gillespie algorithm for the simulation}
\end{algorithm}
\DecMargin{2em}

\subsection*{Simulation running}

The main simulation code is written in C++ (C++17), using the $GDAL$ library for geospatial transformations and manipulations, and the $nanoflann$ k-d tree implementation for radius queries. The population density map from the Global Human Settlement Layer dataset is loaded as a 2D array, with areas inside and outside the border marked separately. The state of the simulation is exported into an output log each day of the simulation as current SEIR counts, and as a PNG file representing the current state of population on the density map where susceptible individuals are represented by grey pixels, exposed — by orange pixels, infected — by red pixels, and removed — by green pixels. Localized infection clusters in the simulation, as well as directions of infection spreading, are therefore clearly visible.

We run a set of 100 simulations, each simulation starting with the same initial condition, in which a set of randomly chosen nodes are infected and all the other $N-n_0$ nodes are susceptible (in this particular configuration, $n_0 = 2$). We measure the percentage of recovered nodes at the end of the simulation (final size), and use the mean estimator from Seaborn \cite{waskom2021seaborn} with $\pm1\sigma$ interval, to display the data from multiple simulations conveniently. The counts of nodes at different states can be aggregated and transformed into different indicators to be displayed as stochastic epidemic curves.

The distance matrix $M$ is constructed using information about population density for each simulated region, obtained from the ESM2015 dataset~\cite{esm2019}. The age-dependent contact matrix $A$ is obtained from the work of Kiesha Prem et al~\cite{prem2017projecting}, which extended the results of the POLYMOD project~\cite{mossong2008social} to 152 countries.

\subsection*{Surrogate model optimization}

To provide the initial epidemic parameters for the full simulation, we use surrogate model optimization. We fit a stochastic SEIR model assuming homogeneous contact network to match the observed data on  COVID-19 cases in Switzerland, provided by ETH Zurich~\cite{foph2021}. 

We wrote an adapter function for running the surrogate model and exporting the simulation results, and compared the results with the actual observed number of hospitalizations for the selected geographical region. Then we run a nonlinear least squares optimization routine \texttt{curve\_fit} from \texttt{scikit-optimize} Python package, which works with arbitrary model functions. 

After the fitting, the determined parameter values are used as inputs to the simulation of the main model, leaving only the local infection distance threshold $r$ and the proportion of global contacts among all contacts as free parameters.

In the first stage of the fitting procedure, we fix the $\epsilon$ and $\gamma$ values of the SEIR model (inverse incubation period and inverse infectiousness period, as these are not changing) as $X$ and $Y$, respectively, and assume that $\beta$ (infectiousness rate) is changing stepwise over time due to the different epidemic control regimes. We assumed that the first infections were introduced between January 15th and February 25th: the first case was officially registered in the canton of Geneva on February 25th, but the infection was likely already spreading at that time. The epidemic control regime was assumed to have changed five times: around March 18th when the first lockdown was introduced, around May 25th when it was lifted, during the the second lockdown in the canton of Geneva (which started November 1st and ended November 29th), and during the second federal lockdown (which started January 18th 2021, currently ongoing as of 19th February 2021).

The following parameters were fitted:
\begin{enumerate}
\item $t_k$ — the regime change dates. Some of the regime change points (the date of introduction of the new pandemic response policies) are known, but it still makes sense to consider them as free parameters: if the optimizer finds a regime change date completely on its own, from data only, it is a part of verification that the model and the optimizer work correctly.
\item $\beta_k$ — infection spreading rates for each time interval ($\beta_0$ bounded from initial estimations of $R_0$).
\item $q$ — the assumed proportion of the infected population who were hospitalized.
\end{enumerate}

The bounds are shown in Table 2.

The fitting routine consists of solving the optimisation problem of finding the minimum of the function:
\begin{equation}
F(\theta)=\sum_{i=1}^N \rho (f_i(\theta)^2), \notag
\end{equation}
where $\theta = (\theta_1,...,\theta_r)$ is a set of parameters which we want to estimate, $N$ is the number of available data points, $\rho$ is a loss function to reduce the influence of outliers, $f_i(\theta)$ is the $i$-th component of the vector of residuals (in this particular simulation, we are fitting for the number of hospitalizations, as the most objective and readily available statistics, assuming that the proportion of hospitalized patients always stays the same for each age group, at least within the region under consideration).

Given a model function $m(t;\theta)$ and some data points $D=\{(t_I,D_I|i=1, .., N\}$ we defined the vector of residuals as the difference between the model prediction and the data, that is $F_i(\theta)=m(t_i;\theta) -d_i$.

\begin{table}
\begin{center}
\begin{tabular}{ c|p{4cm}|c|p{4cm}} 
 \hline
 Name & Description & Prior range / search bounds & Fitted values\\ 
 \hline
 $\beta_k$ & Infection rates (for different regimes) & 0.01--2.5 & 1.0, 0.7, 0.09, 0.2, 0.2, 0.34, 0.59, 0.3, 0.1, 0.3 \\
 \hline
 $\epsilon$ & Incubation rate & 0.15--0.3 & 0.2 \\
 \hline
 $\gamma$ & Removal rate & 0.1--0.3 & 0.16 \\
 \hline
 $t_k$ & Epidemic regimes change times, days from the start of the epidemic & location dependent & [0], 40,  51, 110, 138,  166,  225,  265, 282, 301 \\
 \hline
\end{tabular}
\caption{Free parameters to be fitted in the surrogate model}
\end{center}
\end{table}

We compared the simulation outputs against the real world observation from the COVID-19 clinical dataset, provided by the Swiss Federal Office of Public Health \cite{foph2021}. The epidemic parameters obtained in the fitting step were submitted to the main simulation code, which was then run 100 times independently. The initial mean contact radius $r$ for the local transmission mode was set to 500 meters, as adapted from \cite{kan2008two}, and modified over time between 50m and 500 m according to changes in mobility as described in \cite{LLC}. Proportions of global contacts were adapted from the same study to range between $0.01\%$ and $1.00\%$, in line with mobility data. We used the number of hospitalizations as a reference output for comparison purposes. We considered hospitalizations to be a subclass of $R$ (removed) compartment, as hospitalized patients normally do not infect anybody else. The proportions of infected individuals in each age group assumed to be hospitalized are shown in Fig.3. Additionally, we compared the cumulative percentage of recovered patients in the simulation with the numbers obtained from seroprevalence studies performed in the canton of Geneva \cite{stringhini2020seroprevalence}. 

The set of epidemic parameters obtained in the surrogate model is shown in Table 2. The simulation reproduced the initial epidemic wave of COVID-19 that happened in March-April 2020 (Fig. 2, 3). 
The simulated number of recovered cases is consistent with seroprevalence measurements performed in the canton of Geneva in April-May 2020 and November-December 2020 (Fig. 4).

\section*{Results and discussion}

We developed a flexible network simulation model for COVID-19 and tested it successfully with data from the canton of Geneva, Switzerland. Our system allows us not only to track and predict the epidemic progression, but also to simulate and evaluate various measures introduced to contain the spread of infection, such as limiting mobility and partial or complete lockdown. The simulation software also makes it possible to model the effectiveness of vaccination and to choose which population groups (based on age or location) should be prioritized in vaccination to accelerate the development of population immunity.

As epidemic response organizations often have limited resources, optimal resource allocation is crucial for developing the most efficient interventions. This includes determining the most vulnerable population groups, to identify epidemic hotspots, and to determine the most impactful dates and locations for introducing non-pharmaceutical interventions. Models without spatial dimension have limited ability to answer some questions about intervention effectiveness and resource allocation \cite{Ball2008, riley2015five}. Our framework employs age-based contact matrices to account for differences in social networks and contact patterns of people within different age groups; it uses detailed population density maps to predict (and visualize) geographical locations of epidemic hotspot clusters, and it employs a robust optimization system to adapt itself for real observations and medical statistics available for the chosen simulation area, enabling operators to choose most impactful response options.

Our approach also has some limitations. Full representation of the population on the level of individuals is computationally much more expensive than traditional compartmental models; while the computational resources required are still modest compared to many problems in computational physics, it makes continuous experiments and parameters optimization time-consuming and often impractical. Our approach requires some non-standard parameters which are hard to calculate from the observed data (such as the proportion of “locally” transmitted infections). 

The addition of some new features that could further improve the simulation and the underlying model. Among these are accounting for seasonality, marking up the geographical representation with zoning (e.g. commercial, residential, business or recreational zones, with corresponding differences in mobility patterns); or accounting for travel across the external borders of the simulated setting (which could be represented as spontaneously generated new infection events). Adding these features, however, constitute the risk of propelling the model well beyond analytical tractability, and reduce it to an empirical numerical simulation limiting our ability to extract novel theoretical results. Additional research is required to prioritize the approaches for the greatest robustness and practical utility.

Thus, despite the fact that this model is computationally more expensive than traditional compartmental models and requires some unconventional parameters, our approach can bring more fidelity in the simulation results. This model provides an efficient way to take into account the age and geographical structure and mobility, without resorting to even more computationally intensive direct simulations of people movement. \\

\section*{Funding}

The project was funded by the Swiss National Science Foundation (grant \textnumero 196270).

Liudmila Rozanova was funded by a grant from the European Open Science Cloud
(EOSC, https://www.eoscsecretariat.eu/). The EOSCsecretariat.eu has received funding from the European Union’s Horizon Programme call H2020-INFRAEOSC-05-2018-2019, grant Agreement number 831644.

Olivia Keiser was funded by a professorship grant from the Swiss National Science Foundation (grant \textnumero 163878).

\printbibliography
\pagebreak
\section*{Appendix A}

\begin{table}[hb]
\begin{center}
\tiny{
\begin{tabular}{ |c|c|c|c|c|c|c|c|c|c|c|c|c|c|c|c|c|} 
 \hline
 Age & 0-5	& 5-10	& 10-15	& 15-20 & 20-25	& 25-30	& 30-35	& 35-40	& 40-45 & 45-50	& 50-55	& 55-60	& 60-65	& 65-70	& 70-75	& 75+\\ 
 \hline
 0-5 &	1.35 &	0.56 &	0.25 &	0.14 &	0.21 &	0.39 &	0.69 &	0.73 &	0.49 &	0.22 &	0.22 &	0.18 &	0.14 &	0.13 &	0.07 &	0.04  \\
 \hline
 5-10 &	0.48 &	5.67 &	0.89 &	0.22 &	0.15 &	0.33 &	0.62 &	0.80 &	0.88 &	0.37 &	0.21 &	0.16 &	0.14 &	0.12 &	0.05 &	0.05 \\
 \hline
 10-15 &	0.18 &	1.74 &	9.11 &	0.81 &	0.29 &	0.25 &	0.43 &	0.68 &	1.11 &	0.64 &	0.34 &	0.15 &	0.09 &	0.10 &	0.07 &	0.07 \\
 \hline
 15-20 &	0.09 &	0.29 &	3.11 &	11.68 &	1.50 &	0.77 &	0.64 &	0.79 &	1.09 &	1.20 &	0.67 &	0.24 &	0.08 &	0.06 &	0.03 &	0.03 \\
 \hline
 20-25 &	0.13 &	0.17 &	0.31 &	2.35 &	3.95 &	1.77 &	1.23 &	1.15 &	1.01 &	1.32 &	0.98 &	0.49 &	0.12 &	0.06 &	0.06 &	0.06 \\
 \hline
 25-30 &	0.35 &	0.24 &	0.24 &	0.89 &	2.07 &	3.66 &	1.91 &	1.52 &	1.36 &	1.17 &	1.21 &	0.68 &	0.21 &	0.08 &	0.04 &	0.04 \\
 \hline
 30-35 &	0.59 &	0.80 &	0.64 &	0.52 &	1.06 &	1.82 &	3.13 &	1.98 &	1.58 &	1.26 &	0.96 &	0.67 &	0.29 &	0.14 &	0.06 &	0.06 \\
 \hline
 35-40 &	0.61 &	0.95 &	0.77 &	0.74 &	0.77 &	1.45 &	1.81 &	3.14 &	2.22 &	1.46 &	1.06 &	0.53 &	0.29 &	0.21 &	0.11 &	0.05 \\
 \hline
 40-45 &	0.35 &	0.80 &	1.03 &	1.20 &	0.97 &	1.34 &	1.73 &	1.99 &	3.19 &	1.89 &	1.34 &	0.47 &	0.22 &	0.16 &	0.11 &	0.06\\
 \hline
 45-50 &	0.19 &	0.58 &	0.79 &	1.86 &	1.00 &	1.12 &	1.39 &	1.59 &	1.79 &	2.57 &	1.37 &	0.60 &	0.20 &	0.13 &	0.11 &	0.12\\
 \hline
 50-55 &	0.17 &	0.64 &	1.11 &	1.53 &	1.15 &	1.48 &	1.38 &	1.38 &	1.91 &	2.15 &	2.37 &	1.00 &	0.30 &	0.16 &	0.10 &	0.12\\
 \hline
 55-60 &	0.30 &	0.66 &	0.79 &	0.94 &	0.75 &	1.27 &	1.35 &	1.07 &	1.29 &	1.13 &	1.33 &	1.68 &	0.52 &	0.27 &	0.12 &	0.11\\
 \hline
 60-65 &	0.31 &	0.33 &	0.26 &	0.44 &	0.38 &	0.61 &	0.82 &	0.79 &	0.63 &	0.56 &	0.51 &	0.71 &	1.23 &	0.49 &	0.26 &	0.12\\
 \hline
 65-70 &	0.23 &	0.35 &	0.28 &	0.17 &	0.27 &	0.42 &	0.73 &	0.68 &	0.65 &	0.43 &	0.44 &	0.58 &	0.63 &	1.29 &	0.33 &	0.17\\
 \hline
 70-75 &	0.10 &	0.27 &	0.31 &	0.31 &	0.17 &	0.29 &	0.33 &	0.53 &	0.66 &	0.52 &	0.41 &	0.35 &	0.67 &	0.72 &	1.00 &	0.32\\
 \hline
 75+ &	0.20 &	0.27 &	0.40 &	0.33 &	0.15 &	0.18 &	0.33 &	0.38 &	0.48 &	0.57 &	0.56 &	0.35 &	0.27 &	0.40 &	0.33 &	0.56\\
 \hline
\end{tabular}
}
\caption{A contact matrix for 16 age groups in Switzerland~\cite{prem2017projecting}}
\end{center}
\end{table}
\pagebreak

\begin{figure}[t]
\begin{multicols}{2}
\includegraphics[width=1.0\columnwidth]{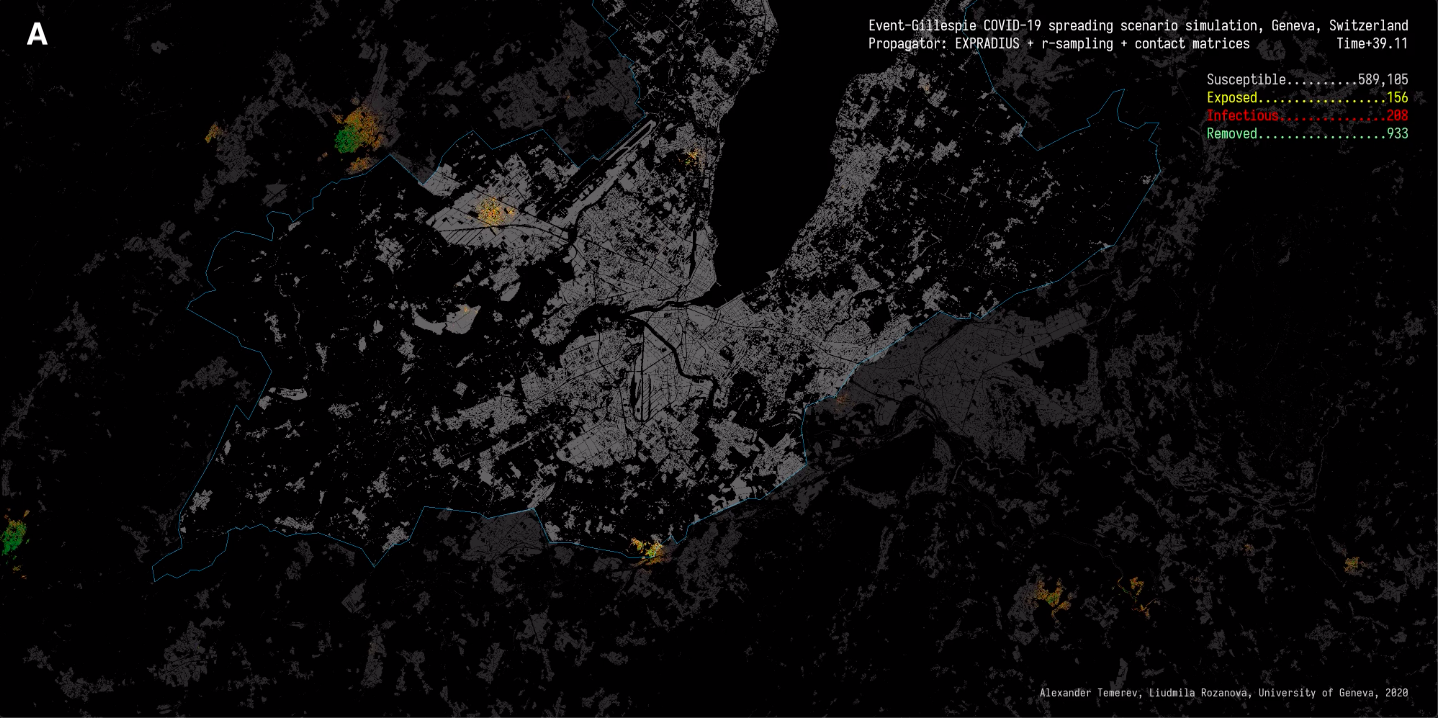}
\includegraphics[width=1.0\columnwidth]{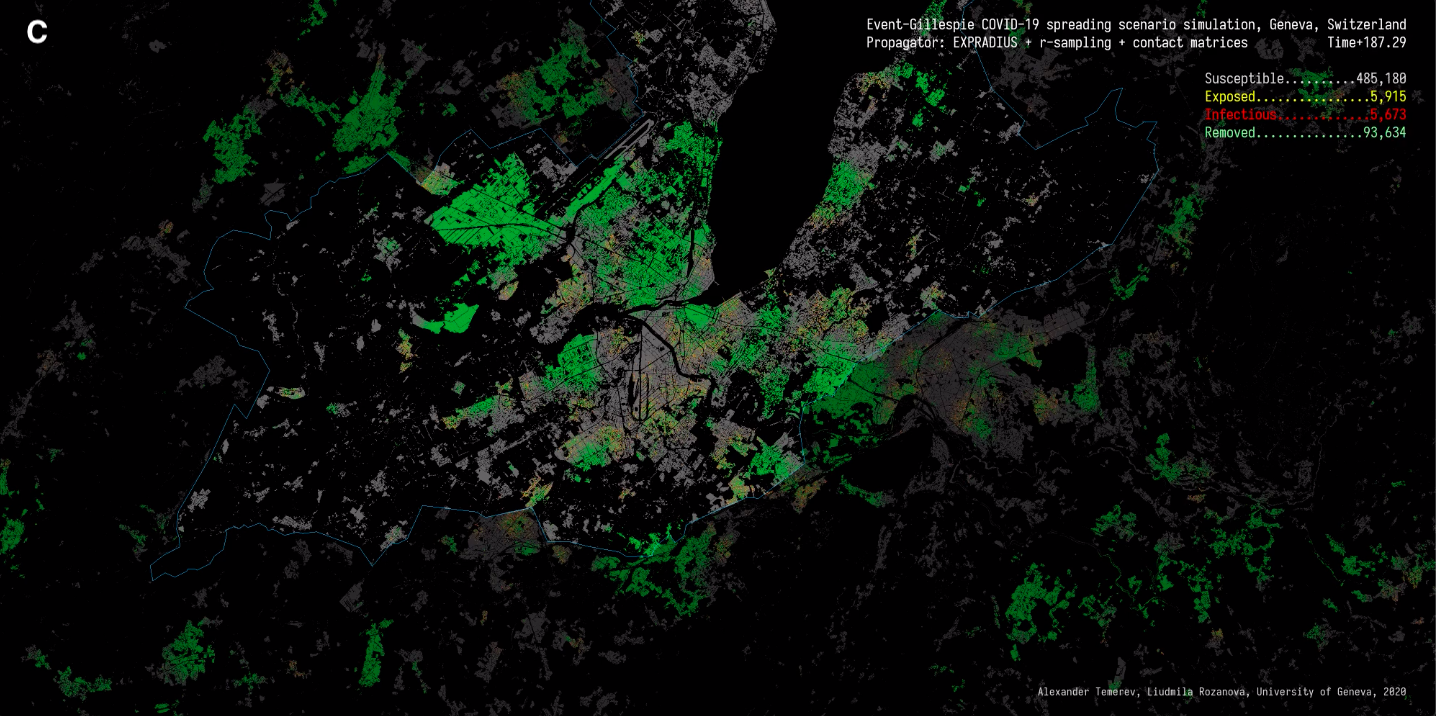}
\includegraphics[width=1.0\columnwidth]{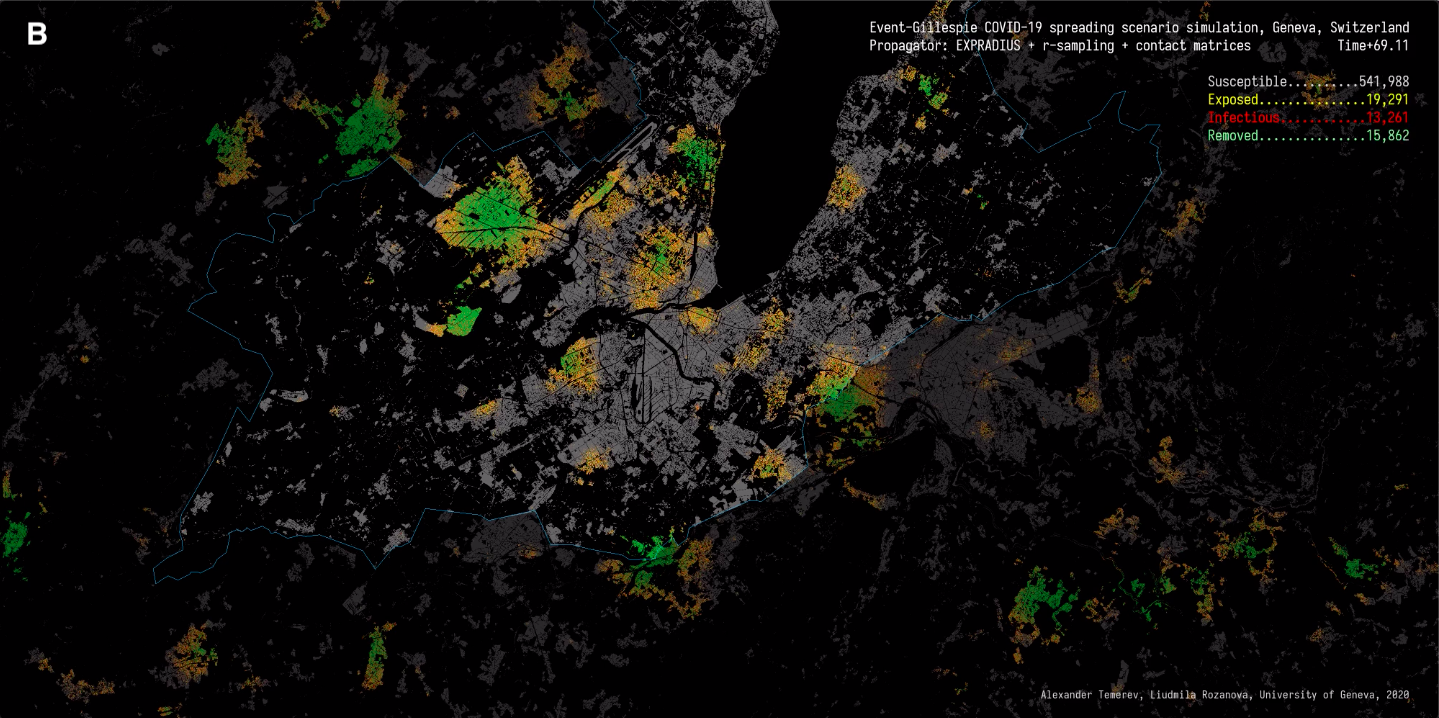}
\includegraphics[width=1.0\columnwidth]{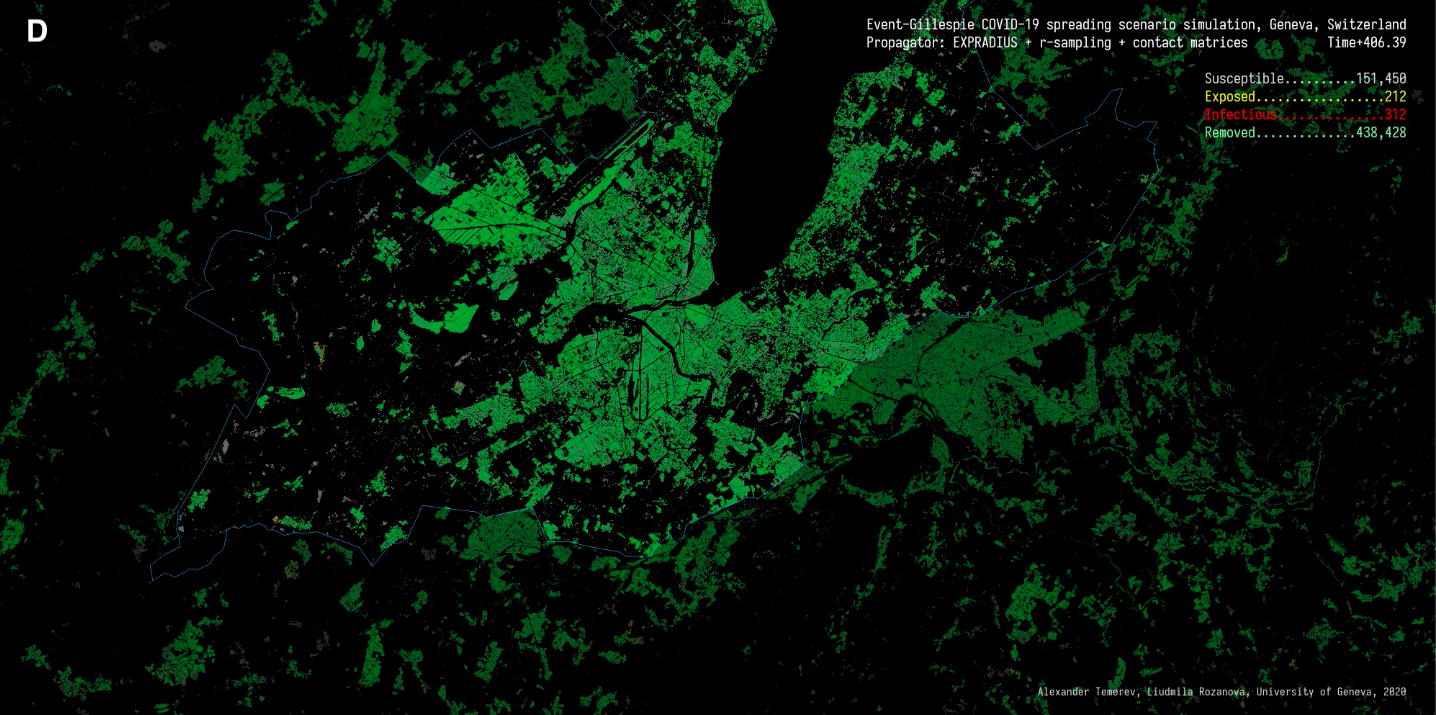}
\end{multicols}
\centering
\caption{Simulation state for the canton of Geneva, Switzerland: A) March 3, 2020; B) April 2, 2020; C) October 6, 2020; D) October 6, 2021.}
\end{figure}

\begin{figure}[t]
\includegraphics[width=1.0\columnwidth]{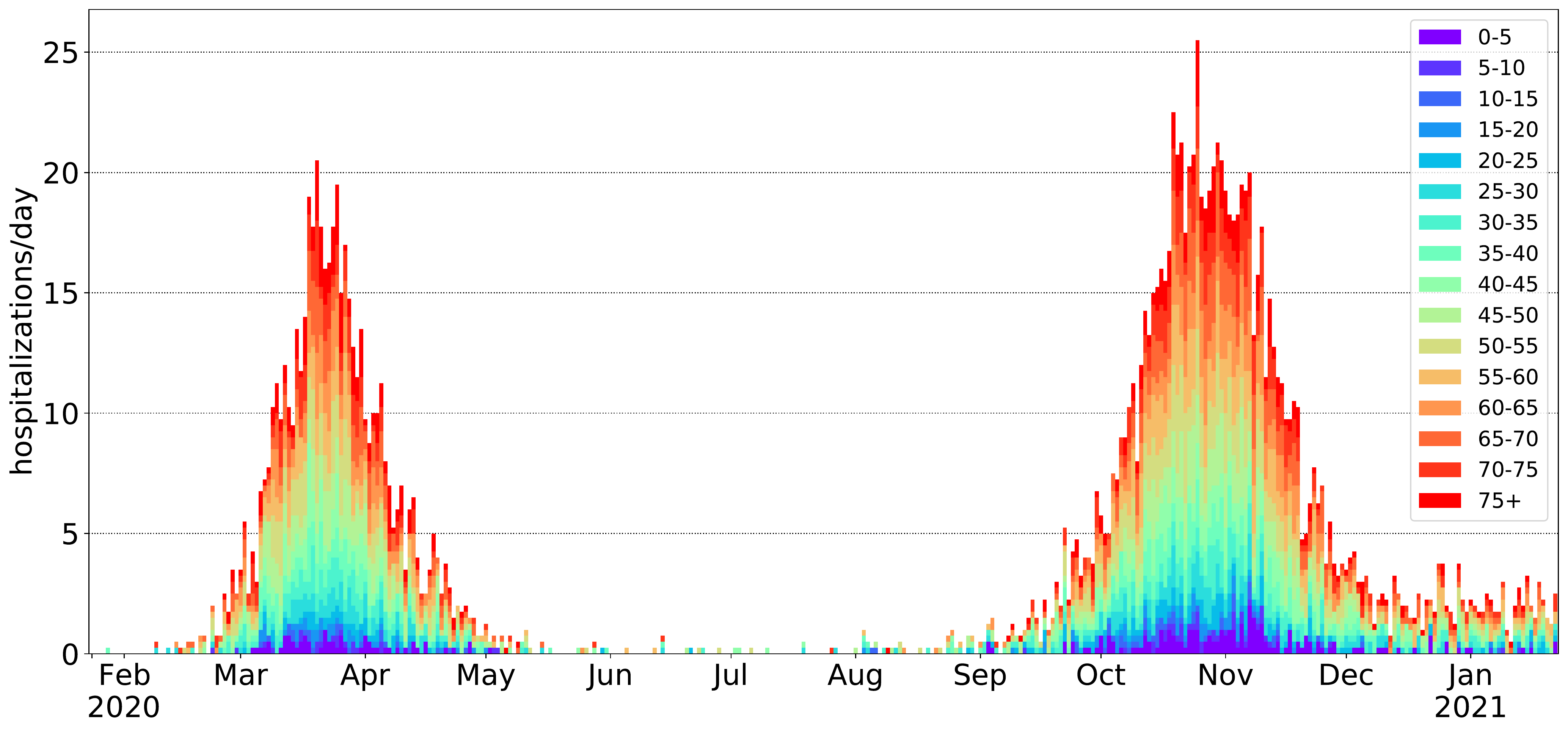}
\centering
\caption{Simulation: daily hospitalizations in the canton of Geneva, by age groups}
\end{figure}

\begin{figure}[t]
\includegraphics[width=1.0\columnwidth]{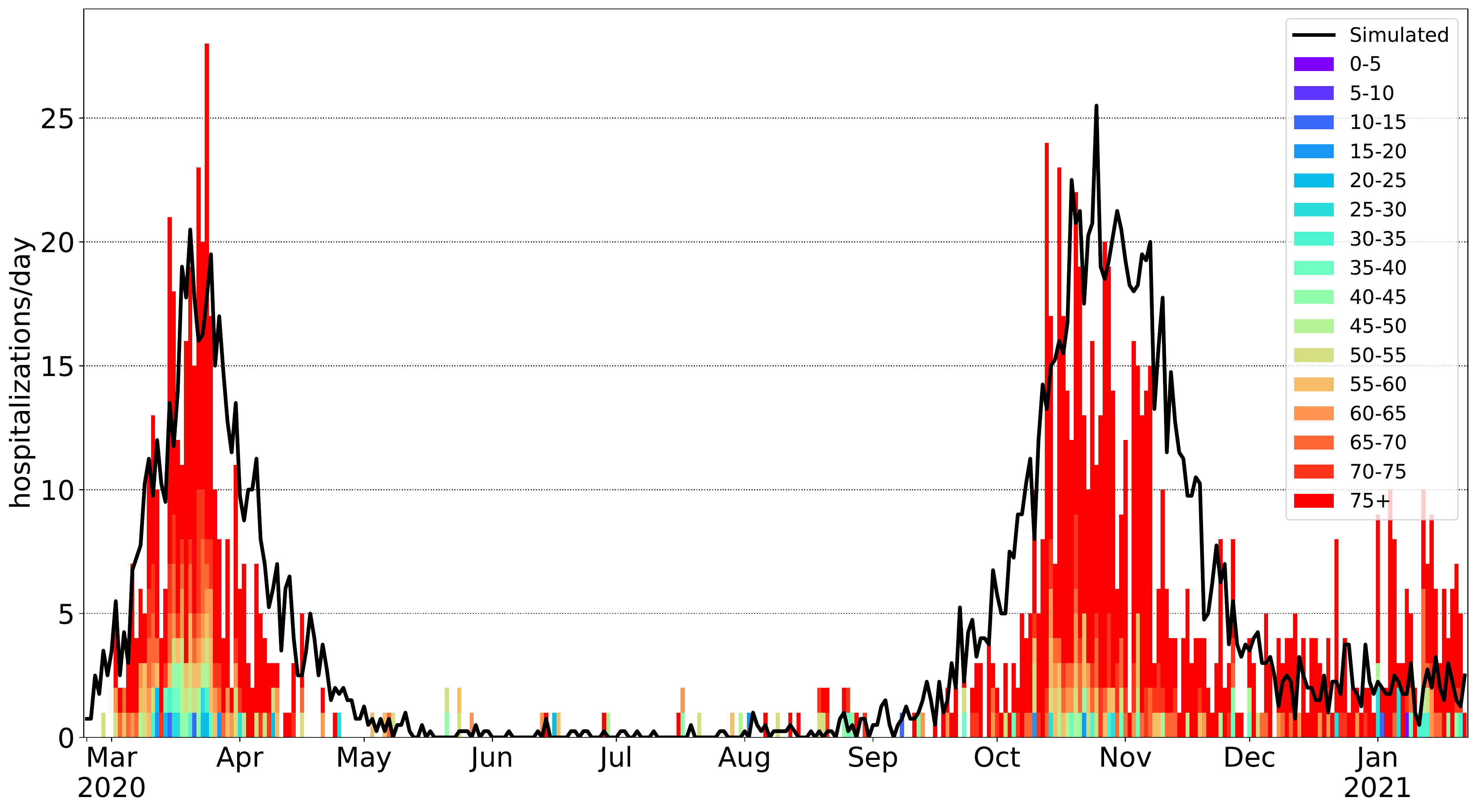}
\centering
\caption{Daily hospitalizations in Geneva canton, by age groups, compared to the simulation results}
\end{figure}

\begin{figure}[t]
\includegraphics[width=1.0\columnwidth]{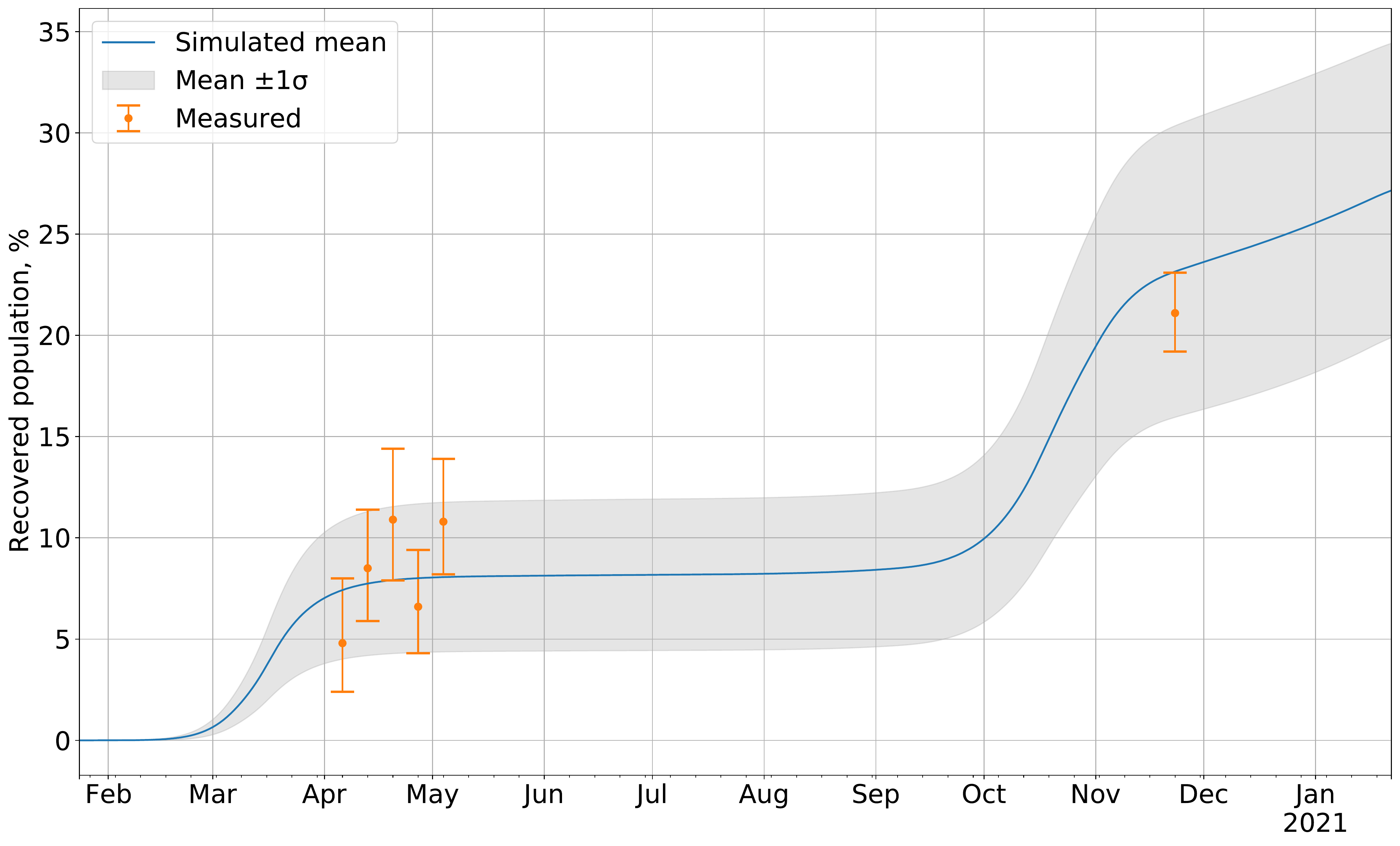}
\centering
\caption{Simulation: percentage of recovered population in the canton of Geneva, aggregated data from 100 experiments with $\pm1\sigma$ error bands, compared to measured seroprevalence values}
\end{figure}

\end{document}